\documentclass[%
 reprint,
superscriptaddress,
 amsmath,amssymb,
 aps,
]{revtex4-2}

\usepackage{graphicx}% Include figure files
\usepackage{dcolumn}% Align table columns on decimal point
\usepackage{bm}% bold math
\usepackage{braket}
\begin{document}

\preprint{APS/123-QED}

\title{Wide range linear magnetometer based on a sub-microsized K vapor cell}

\author{M. Auzinsh}
\affiliation{Department of Physics, University of Latvia, Rainis boulevard 19, LV-1586 Riga, Latvia}
\author{A. Sargsyan}
\affiliation{Institute for Physical Research, NAS of Armenia, Ashtarak-2, 0203 Armenia}
\author{A. Tonoyan}
\affiliation{Institute for Physical Research, NAS of Armenia, Ashtarak-2, 0203 Armenia}
\author{C. Leroy}
\affiliation{Laboratoire Interdisciplinaire Carnot de Bourgogne, UMR CNRS 6303, Université Bourgogne Franche-Comté, 21000 Dijon, France}
\author{R. Momier}
\email{rodolphe.momier@u-bourgogne.fr}
\affiliation{Institute for Physical Research, NAS of Armenia, Ashtarak-2, 0203 Armenia}
\affiliation{Laboratoire Interdisciplinaire Carnot de Bourgogne, UMR CNRS 6303, Université Bourgogne Franche-Comté, 21000 Dijon, France}
\author{D. Sarkisyan}
\affiliation{Institute for Physical Research, NAS of Armenia, Ashtarak-2, 0203 Armenia}
\author{A. Papoyan}
\affiliation{Institute for Physical Research, NAS of Armenia, Ashtarak-2, 0203 Armenia}

\date{\today}% It is always \today, today,

\begin{abstract}  $^{39}$K atoms have the smallest ground state ($^2S_{1/2}$) hyperfine splitting of all the most naturally abundent alkali isotopes and, consequently, the smallest characteristic magnetic field value $B_0 = A_{^2S_{1/2}}/\mu_B \approx 170$ G, where $A_{^2S_{1/2}}$ is the ground state's magnetic dipole interaction constant. In the hyperfine Paschen-Back regime ($B \gg B_0$, where $B$ is the magnitude of the external magnetic field applied on the atoms), only 8 Zeeman transitions are visible in the absorption spectrum of the $D_1$ line of $^{39}$K, while the probabilities of the remaining 16 Zeeman transitions tend to zero. In the case of $^{39}$K, this behavior is reached already at relatively low magnetic field $B > B_0$. For each circular polarization ($\sigma^-,\sigma^+$), 4 spectrally resolved atomic transitions having a sub-Doppler width are recorded using a sub-microsized vapor cell of thickness $L = 120 - 390$ nm. We present a method that allows to measure the magnetic field in the range $0.1 - 10$ kG with micrometer spatial resolution, which is relevant in particular for the determination of magnetic fields with a large gradient (up to 3 G$/\mu$m). The theoretical model describes well the experimental results.
\end{abstract}

\maketitle

\section{Introduction}

Strong magnetic fields ($0.1-10$ kG) are necessary for the application of magneto-optical processes using alkali vapors \cite{RevModPhys74.1153,OPA,PhysRevA84.063410,ApplOpt59.2231}. In strong magnetic fields, which are determined from the conditions $B \gg B_0 = A_{\text{hfs}}/\mu_B$ where $A_{\text{hfs}}$ is the magnetic dipole interaction constant of the state $^2S_{1/2}$ and $\mu_B$ is the Bohr magneton (the constants are given in \cite{CompPhysComm189.162,CODATA,RevModPhys.49.31}), the total angular momentum $J$ and nuclear spin $I$ are decoupled \cite{PhysRevA84.063410}. This is called hyperfine Paschen-Back (HPB) regime \cite{HPB,PhysRevA84.063410,OptLett37.1379}. In this regime the energy levels are described by the magnetic numbers $m_J$ and $m_I$. The $B_0$ values for $^{85}$Rb, $^{87}$Rb and $^{133}$Cs are approximately 700, 2400 and 1700 G, respectively. For $B \geq 2 B_0$, the number of atomic transitions in the absorption spectrum decreases significantly. In \cite{OptLett37.3405}, an optical isolator based on a Rb vapor was described using $B \sim 6$ kG. In \cite{PhysRevA92.063810}, interesting features were found in the Saturated Absorption (SA) spectrum of a Rb vapor in strong magnetic field at the cross-over resonances. In \cite{JModOpt65.713}, a 4-wave mixing process was carried out in a $2$ mm-size Rb cell at $B \sim 6$ kG. It was shown in \cite{OptComm284.4007} that at $B > 4$ kG, only 10 transitions are present in the transmission spectrum of the Rb $D_1$ line. In \cite{OptLett39.2270}, a $40$ $\mu$m-size Rb cell was used, in which the SA process was studied at $B \sim 6$ kG and the laser frequency was stabilized on low frequency-shifted atomic transitions.

For the formation of a strong magnetic field, permanent magnets made of a neodymium-iron-boron alloy are used, which create a magnetic field $B \sim 4$ kG near their surface, with a gradient of $\sim 0.15$ G$/\mu$m. In \cite{OptLett42.1476}, Selective Reflection (SR) of laser radiation from a nanocell (NC) containing a Rb vapor was used to measure magnetic fields $> 1000$ G. However, due to the large value of $B_0({}^{87}\text{Rb}) = 2400$ G the SR method was not convenient to use for measurement of magnetic fields $< 1000$ G, since the Zeeman components of $^{87}$Rb and $^{85}$Rb remain partially overlapped. Therefore, the development of methods for recording homogeneous and inhomogeneous magnetic fields in a wide range is important.

%%CORRECTED 13:16 15 MAR 22

It is promising to use $^{39}$K for which $B_0({}^{39}\text{K}) = 170$ G is small \cite{EPL110.23001,JETP153.355} and for a $B$-field $>170$ G, only $8$ Zeeman transitions remain in the transmission spectrum of $^{39}$K $D_1$ line ($4$ transitions for each circular polarization $\sigma^\pm$). In this work, these transitions are used to measure magnetic fields with a high spatial resolution.

It is important to note that using transmission is more convenient than using SR since the resonances in the transmission spectrum have a symmetrical shape, whereas the SR technique gives rise to dispersion-like line shapes which are also sensitive to the nanocell thickness around $L = \lambda/2$ \cite{JETPLett104.224,JOSAB20.793,PhysRevA.51.1959}.

\section{Theoretical model: $^{39}$K $D_1$ line Zeeman transitions intensities within an external magnetic field}

\begin{figure*}{h!}
\centering
\includegraphics[scale=1]{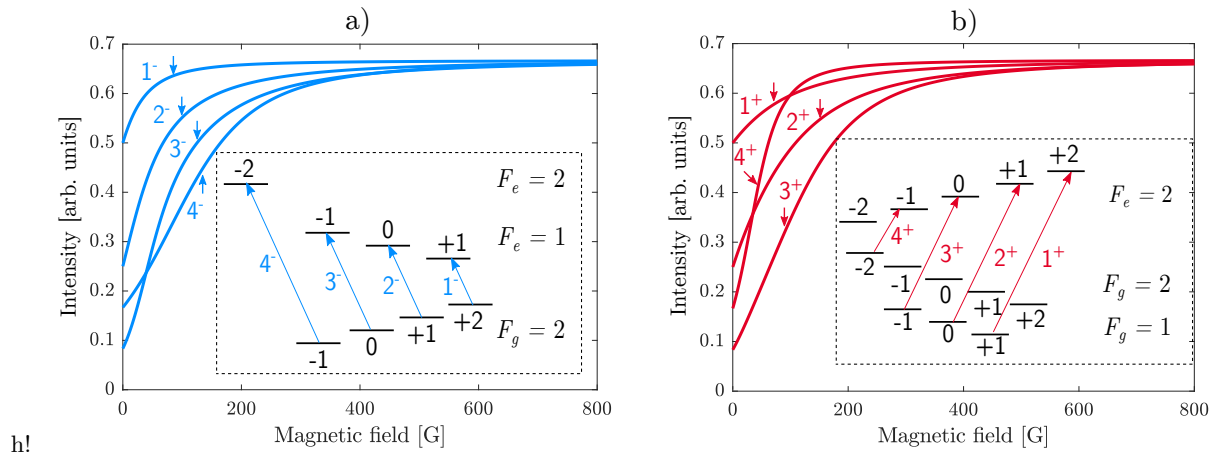}
\caption{Magnetic field dependence of Zeeman transition intensities of the $D_1$ line of $^{39}$K. a) $\sigma^-$ transitions.  b) $\sigma^+$ transitions. Transition labels are described in the insets. Transitions having a probability close to zero for $B > 100$ G are omitted.\label{fig:1}}
\end{figure*}

In Fig.~\ref{fig:1} we present the theoretical magnetic field dependence of the $^{39}$K $D_1$ line Zeeman transitions intensities (probabilities) for $\sigma^+$- and $\sigma^-$-polarized laser radiation. For a relatively small magnetic field $B > 100$ G, there are in total $24$ atomic transitions for $\sigma^\pm$ radiations, meanwhile for $B > 2B_0({}^{39}$K$)$ only $8$ transitions with approximately the same amplitudes and equidistant in frequency remain in the transmission/absorption spectra when interrogating the vapor with circularly polarized laser radiation (with respect to the magnetic field direction), meaning HPB regime is reached. The theory describing the modifications of the Zeeman transitions probabilities in a static magnetic field is described in details in \cite{EPL110.23001,PhysRevA42.2766,JQSRT272.107780}.

Hereafter, we recall the main ingredients of the theoretical model. Under the influence of an external magnetic field, the atomic hyperfine states $F$ split into Zeeman sublevels $\ket{F,m_F}$. Rigorously, the magnetic field gives rise to mixed states $\ket{\psi(F,m_F)}$ such that 
\begin{equation}
\ket{\psi(F,m_F)} = \sum_{F'} \alpha_{F,F'}(B)\ket{F',m_F}\label{eq:mix}
\end{equation} 
where only the states having the same $m_F$ are coupled due to the selection rules on the matrix elements \cite{PhysRevA42.2766}. The intensity (probability) $A_{eg}$  of a transition between two Zeeman states is proportional to the so-called "modified transfer coefficient"
\begin{equation}
A_{eg} \propto a^2[\psi(F_e,m_{F_e});\psi(F_g,m_{F_g});q]\, .
\end{equation}
The latter are defined by 
\begin{align}
&a[\psi(F_e,m_{F_e});\psi(F_g,m_{F_g});q] \nonumber\\
&= \sum_{F_g',F_e'} \alpha_{F_e,F_e'} a(F_e',m_{F_e};F_g',m_{F_g},q)\alpha_{F_g,F_g'}\, ,
\end{align}
where $\alpha_{F,F'}$ are the mixing coefficients described by equation \eqref{eq:mix}, obtained by numerical diagonalization of the Zeeman Hamiltonian \cite{PhysRevA42.2766}. The indices $g$ and $e$ stand for ground and excited states throughout the whole paper.
The so-called "unperturbed" transfer coefficients $a(F_e,m_{F_e};F_g,m_{F_g},q)$ are given by
\begin{align}
	&a(F_e,m_{F_e};F_g,m_{F_g},q) = (-1)^{1+I+J_e+F_e+F_g-m_{F_e}}\sqrt{2F_e+1}\nonumber \\
	&\times\sqrt{2F_g+1}\sqrt{2J_e+1}\begin{pmatrix}F_e & 1 & F_g \\ -m_{F_e} & q & m_{F_g}\end{pmatrix}\begin{Bmatrix}F_e & 1 & F_g \\ J_g & I & J_e\end{Bmatrix}\, .
\end{align}
where $J_g = J_e = 1/2$ and $I = 3/2$ for the $D_1$ line of $^{39}$K. These values lead to $F_g = 1,2$ and $F_e = 1,2$. The index $q$ reflects the polarization of the laser: $q=0,\pm 1$ for $\pi,\sigma^\pm$ polarization respectively. The parentheses (resp. curly brackets) denote Wigner $3j$- (resp. $6j$-) symbols.
Due to its thickness, the NC used in our experiments can be seen as a Fabry-P\'{e}rot interferometer of length $L$ (quality factor $Q = 1-r_w^2\exp(2ikL)$) filled with an atomic vapor. As described in \cite{JOSAB20.793}, the transmitted signal can be expressed as 
\begin{equation}
S_t \simeq 2t_{wc}t_{cw}^2E_i \mathrm{Re}\left\lbrace I_f - r_wI_b\right\rbrace/|Q|^2\, ,
\end{equation}
where $t$ and $r$ are transmission and reflection coefficients ($c$ stands for cell and $w$ for window). The quantities $I_f$ and $I_b$ are the forward and backward integrals of the atomic response, which expressions are detailed in \cite{JOSAB20.793}. This model, looped over all Zeeman transition frequencies allows to compute absorption (or transmission) spectra with sub-Doppler resolution, each resonance having an amplitude mainly proportional to $A_{eg}$.

%% CORRECTED 13:48 15 MAR 22 

\section{Experiment}
\subsection{Spectroscopic nanocell filled with $^{39}$K vapor}

A specially fabricated nanocell (NC) has a relatively large area of thickness in the range of $50-400$ nm \cite{EPL110.23001}. A photograph of the NC is shown in Fig.~1 in \cite{EPL110.23001}. The NC windows are $1.5$ mm-thick and their surface is $20\times30$ mm$^2$. They were made of well-polished crystalline sapphire with a roughness around $3$ nm. To minimize the birefringence effects, the $C$-axis was oriented perpendicular to the window surface. A sapphire extension (reservoir) was filled with metallic (natural) $K$ ($93.25\%$ $^{39}$K, $0.01\%$ $^{40}$K and $6.7\%$ $^{41}$K) and heated to a temperature of $140~^\circ$C during the experiment, which allowed to reach the atomic concentration of $N\sim 7\times 10^{12}$ cm$^{-3}$ (a detailed description of the NC was given in \cite{EPL110.23001}).

\subsection{Experimental setup}
\begin{figure*}[ht]
\centering
\includegraphics[scale=0.22]{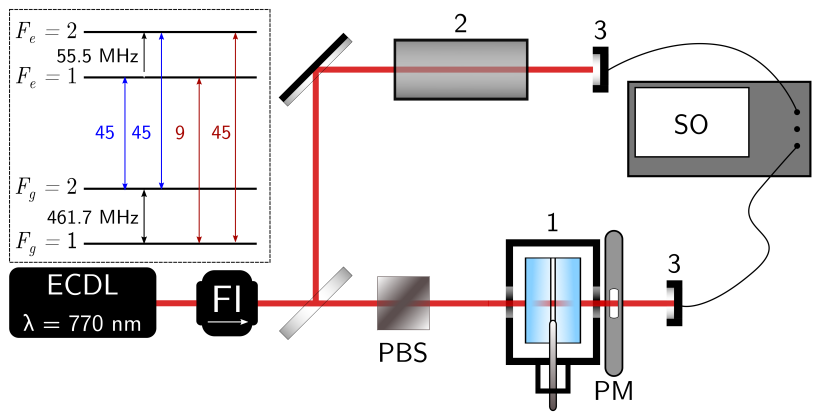}
\caption{Scheme of the experimental setup. ECDL: extended-cavity diode laser ($\lambda = 770$ nm). FI: Faraday isolator. 1: nanocell (NC) filled with $K$ atoms. The thickness $L$ of the vapor columns can be adjusted in the range $120-390$ nm. PBS: polarizing beam splitter. 2: auxiliary cm-long cell filled with K used to form a reference SA spectrum. PM: permanent magnet. $\bold B$ is oriented along the laser propagation direction ($\bold B \equiv B_z, \bold k \equiv k_z$). 3: photo-detectors. SO: oscilloscope. \label{fig:2}}
\end{figure*}

The experimental layout is depicted in Fig.~\ref{fig:2}.  A VitaWave external-cavity diode laser with a wavelength of $\lambda = 770$ nm and a spectral linewidth of $\sim 1$ MHz \cite{RevSciInstrum77.013102} was used. As described earlier, a magnetic field was formed using a strong permanent magnet (PM) placed near the rear NC window and calibrated using an HT201 Teslameter magnetometer. The magnetic field was varied by changing the distance between the PM and the window. To form a frequency reference, a part of the laser radiation was fed to a cm-size K cell  in order to record zero-field Saturated Absorption (SA) spectra \cite{LasPhys6.670,JOSAB2.1431,AJP64.1432}.  The signals were recorded by FD-24K photodiodes (3),  then amplified and fed to a Tektronix digital oscilloscope. 
 	
\subsection{Experimental results}

As a first step, the nanocell (NC) thickness in the direction of laser beam propagation was chosen equal to half the resonant wavelength ($L =\lambda/2= 385$ nm). The technique for measuring the thickness of an atomic vapor column in the NC is described in \cite{AppPhysB71.561,LasFun}. It has been demonstrated earlier that in this case (so called  $\lambda/2$ method), narrowing of atomic transitions (lines) in the absorption spectrum $A(\omega)$ of the NC occurs \cite{JOSAB20.793,EPL110.23001,OptLett44.5533}. This effect is called Coherent Dicke Narrowing \cite{PhysRev.99.532}. To obtain further narrowing of the atomic lines, we performed second derivative (SD) of the absorption spectrum $A''(\omega)$ \cite{OptLett44.5533,PhysLettA390.127114,JOSAB39.973,PhysLettA434.128043}.
	
\begin{figure*}[ht]
\centering
\includegraphics[scale=0.65]{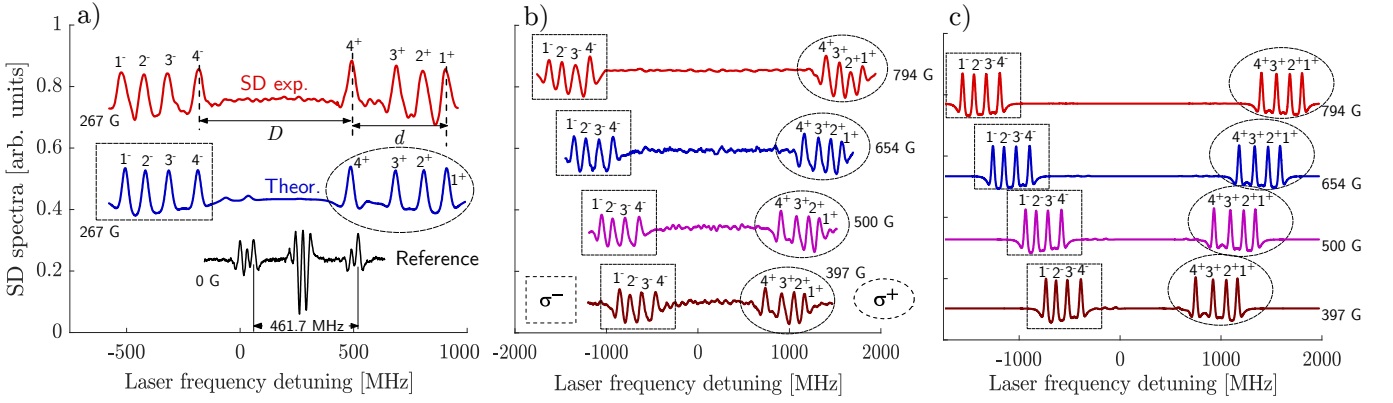}
\caption{$^{39}$K $D_1$ line spectra recorded for $L = 385$ nm. a) Upper curve: experimental SD absorption spectrum for $\sigma^+$ and $\sigma^-$ radiation recorded for $B = 267$ G. Middle curve: theoretical SD absorption spectrum. Lower curve: SD of a saturated absorption spectrum for reference. b) Experimental SD absorption spectra for $B$ increasing from $397$ to $794$ G. The transition labelling in consistent with \ref{fig:1}, $\sigma^-$ and $\sigma^+$ transitions are shown in the rectangle and oval boxes respectively. c) Theoretical SD absorption for the values of $B$ shown in panel b).\label{fig:3}}
\end{figure*}

	In Fig.~\ref{fig:3}a, the upper curve shows an experimental SD absorption spectrum of transitions $F_g=1,2 \rightarrow F_e=1,2$ of the $D_1$ line of $^{39}$K for linearly polarized laser radiation (consisting of $\sigma^+$ and $\sigma^-$ radiations) and  a longitudinal $B$-field of $265$ G. As mentioned earlier, the NC thickness  is $L=385$ nm, the reservoir temperature  is $140~^\circ$C,  and the laser power  is 30 $\mu$W. There are four transitions that are  excited by $\sigma^-$ radiation located on the low-frequency wing of the spectrum, while the four transitions that are  excited by $\sigma^+$ radiation are located on the high-frequency wing of the spectrum.  The resonances are separated by nearly the same frequency interval $\sim 150$ MHz, which  also remains the same for $B\gg B_0$. The middle curve shows  a calculated SD absorption spectrum which is well consistent with the experimental curve. The lower curve shows  the second derivative of a SA spectrum obtained  with a cm-size K cell. In Fig.~\ref{fig:3}b,  experimental SD absorption spectra are shown for $B$ increaseding from $397$ G to $794$ G. As before, there are four transitions (shown in the dashed boxes) that are excited by $\sigma^-$ radiation and located on the low-frequency wing of the spectrum, while the four transitions that are produced by $\sigma^+$ radiation shown in the dashed ovals are located on the high-frequency wing of the spectrum. Fig.~\ref{fig:3}c shows calculated SD absorption spectra (FWHM is chosen to be 40 MHz) for the same conditions as in Fig.~\ref{fig:3}b. The inaccuracy in the determination of strong B-fields is $5\%$, arising from the inhomogeneity of the magnetic field and the mm-dimensions of the measurement head of the HT201Teslameter. One can observe that by measuring the frequency distance $D$ between transitions labelled $4^+$ and $4^-$ obtained by $\sigma^+$ and $\sigma^-$ radiations and using the reference frequency interval of $461.7$ MHz (obtained by SA, see Fig.~\ref{fig:3}a), it is possible to determine the magnitude of the $B$-field using the curve shown in Fig.~4a. Note that at $B= 10$ kG, we measure $D = 37$ GHz.

The narrow-band distributed feedback (DFB) diode-laser described in \cite{PhysScript95.015404} has a linear frequency scanning range of $\sim 40$ GHz. Such type of laser could thus be used for this method of $B$-field measurement in the range $0.1-10$ kG. The ECDL we used has a small linear frequency scanning range of $\sim 4$ GHz, making it only suitable for $B$-field measurements below $2.5$ kG. This range of measurement for the determination of magnetic fields could be increased greatly with the use a DFB laser. In \cite{OptLett37.1379} (Fig.~2) we described a system that allows one to create a $10$ kG magnetic field in a nanocell.

\begin{figure*}[ht]
\centering
\includegraphics[scale=0.65]{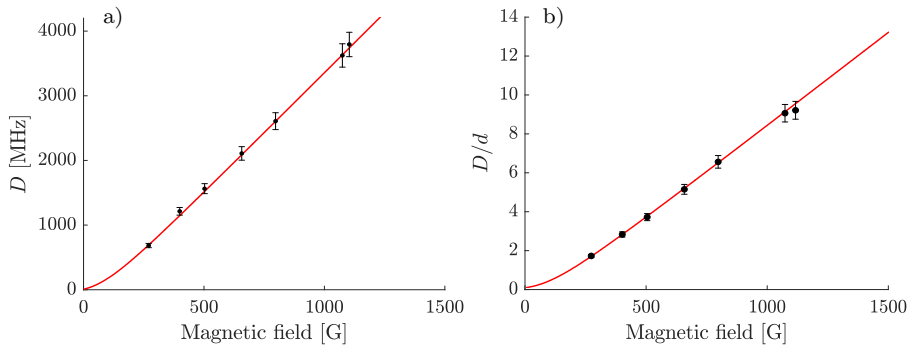}
\caption{a) Frequency difference $D$ between transitions $4^+$ and $4^-$ as a function of the magnetic field. Solid line: theory. Dots with error bars: experimental measurements, the inaccuracy is 5$\%$. b) $D/d$ as a function of the magnetic field. Solid line: theory. Dots with error bars: experimental measurements, the inaccuracy is 5$\%$. \label{fig:4}}
\end{figure*}

It should be noted that magnetic field measurements can also be carried out without using the frequency reference. This is made possible by measuring $D$ and dividing it by the frequency distance $d$ between atomic transitions $4^+$ and $1^+$ (shown in Fig.~\ref{fig:3}a). The relationship $D/d$ as a function of magnetic field is shown in Fig.~\ref{fig:4}b. 
The intervals $D$ and $d$ between atomic transitions are easily determined manually. A specially written computer program (CP) substantially speeds up the data processing. Such program can be written in any language. It is intended for calculating $B$ with the aid of a spectral analysis done on a computer (for this, the data displayed on the oscilloscope is quickly transferred to a computer). The CP finds the significant maxima of the spectrum. The atomic transition with number $1^+$, on the high-frequency wing, is singled out as the first one.
Then CP highlights, in descending order of coordinates, the fourth ($4^+$) and fifth ($4^-$) maxima, determines the distances (along the frequency axis) between the first and fourth maxima ($d$) and fourth and fifth maxima ($D$), and then calculates the ratio $D/d$. Further, to find $B$, the file of the dependence of the ratio $D/d$ with respect to $B$ is used (Fig.\ref{fig:4}b).

\begin{figure*}[ht]
\centering
\includegraphics[scale=0.8]{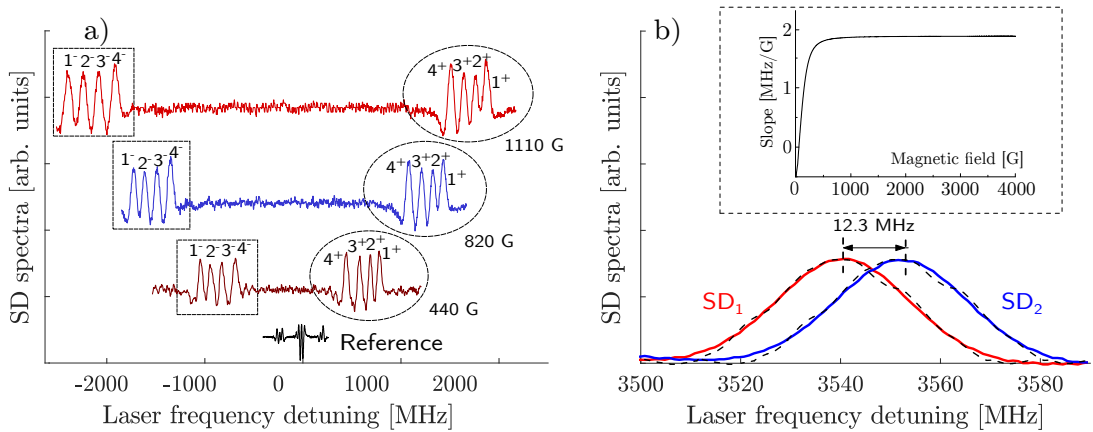}
\caption{$^{39}$K $D_1$ line spectra recorded for $L = 120\pm 5$ nm. a) SD absorption spectra for $B$ increasing from $440$ to $1110$ G. Lower curve: SD of a saturated absorption spectrum for reference. Transition labelling is consistent with Fig.~\ref{fig:1}, $\sigma^-$ and $\sigma^+$ are shown in rectangle and oval boxes respectively. b) Theoretical (black dashed lines) and experimental (red and blue lines) SD spectra for $B_1 = 2$ kG and $B_2 = 2.007$ kG respectively, used to measure $3.3$ G/$\mu$m gradient when the NC ($L=120\pm 5$ nm) is placed on a micrometer stage at $B_1 = 2000$ G and shifted by 2 $\mu$m relative to the initial position when $B_2$ becomes $2007$ G. As can be seen, when moving the NC by 2 $\mu$m, the frequency shift is $12.3$ MHz, which can be measured well. The inset shown the slope [MHz/G] of the $F_g = 1,2 \rightarrow F_e = 1,2$ transitions.\label{fig:5}}
\end{figure*}

For some applications it is important to record absorption spectra at smaller thickness $L$. In Fig.~\ref{fig:5}a, experimental SD absorption spectra are presented for $B$ increasing from $440$ G to $1110$ G. The NC thickness is $L=120 \pm 5$ nm and the reservoir temperature is $155~^\circ$C. The laser power is $30$ $\mu$W. There are still eight well-resolved transitions that are excited by circularly polarized radiation. 
	
Apparently, $L=120 \pm 5$ nm is the minimum acceptable NC thickness since, as shown in \cite{OptLett42.1476}, for $L < 100$ nm a rapid broadening of the resonance lines occurs.
This is due to atom-surface interaction between the vapor and the NC windows (so-called van der Waals interaction), which leads both to an asymmetric broadening of the resonance lines and to their frequency shift towards the low-frequency region of the spectrum \cite{BLOCH200591,PhysRevLett.112.253201}.

The paper \cite{RevSciInstrum91.053202} provides a description of a Stern-Gerlach type deflecting magnet, intended to deflect beams of paramagnetic nanoclusters, molecules and atoms using a magnetic field with a gradient of 3.3 G/$\mu$m.  Figure \ref{fig:5}b) shows theoretical SD absorption spectra (red and blue solid lines with FWHM $80$ MHz and fitted with a Voigt profile) and experimental SD absorption spectra (presented by black dotted lines) that can be used to measure such a gradient if a NC with a thickness $L=120 \pm 5$ nm is placed on a micrometer stage (magnetic field $B_1=2$ kG) and shifted by $2$ $\mu$m with respect to the initial position (the magnetic field becomes $B_2 = 2006.6$ G). The red curve has been calculated for $B_1 = 2000$ G, and the blue curve has been calculated for $B_2=2006.6$ G. Black curves depicts an experimental SD absorption spectrum for $B_1$ and $B_2$. Transition $4^+$ was chosen for measurements. As can be seen from Fig.~\ref{fig:5}b, when moving the NC by 2 microns (fields $B_1$ and $B_2$), the frequency shift is $12.3$ MHz which can be measured well. Note that if the gradient is $2$ times larger, then by moving the NC by 1 micron the frequency shift will also be $12.3$ MHz. It is crucial to note that the use of the above-mentioned Teslameter HT201, of which the sensor part has an area of a few mm$^2$, will lead to large inaccuracies in the determination of magnetic fields with a large gradient. The inset in Fig.~\ref{fig:5}b shows the slope (MHz/G) of $F_g=1,2 \rightarrow F_e=1,2$ transitions, which is $\sim 1.9$ MHz/G for $B >1000$ G and remains nearly constant at higher $B$-field.
               
\section{Conclusion}

The paper demonstrates that the small value $B_0=170$ G of $^{39}$K $D_1$ line (the smallest among all the most naturally abundent alkali metal isotopes) and the use of a sub-microsized vapor cell (also called a nanocell) is convenient for high spectral resolution table-top magnetometry since, for $B>B_0$, only $8$ atomic transitions among the  $24$ initial transitions remain in the absorption spectrum of linearly polarized radiation (in a longitudinal magnetic field). This is a manifestation of the HPB regime.
Two methods are given for the determination of the magnitudes of magnetic fields by recording absorption spectra of atomic vapors having a thickness ranging between $120$ and $390$ nm, which makes it possible to determine a wide range ($0.1 - 10$ kG) of both uniform and strongly inhomogeneous magnetic fields. It has been demonstrated that due to the small thickness $L=120\pm 5$ nm of the $^{39}$K vapor column, large gradients of magnetic fields can also be determined. Such fields are used in Stern-Gerlach-type experiments.

\bibliography{biblio}% Produces the bibliography via BibTeX.

\begin{thebibliography}{37}%
\makeatletter
\providecommand \@ifxundefined [1]{%
 \@ifx{#1\undefined}
}%
\providecommand \@ifnum [1]{%
 \ifnum #1\expandafter \@firstoftwo
 \else \expandafter \@secondoftwo
 \fi
}%
\providecommand \@ifx [1]{%
 \ifx #1\expandafter \@firstoftwo
 \else \expandafter \@secondoftwo
 \fi
}%
\providecommand \natexlab [1]{#1}%
\providecommand \enquote  [1]{``#1''}%
\providecommand \bibnamefont  [1]{#1}%
\providecommand \bibfnamefont [1]{#1}%
\providecommand \citenamefont [1]{#1}%
\providecommand \href@noop [0]{\@secondoftwo}%
\providecommand \href [0]{\begingroup \@sanitize@url \@href}%
\providecommand \@href[1]{\@@startlink{#1}\@@href}%
\providecommand \@@href[1]{\endgroup#1\@@endlink}%
\providecommand \@sanitize@url [0]{\catcode `\\12\catcode `\$12\catcode
  `\&12\catcode `\#12\catcode `\^12\catcode `\_12\catcode `\%12\relax}%
\providecommand \@@startlink[1]{}%
\providecommand \@@endlink[0]{}%
\providecommand \url  [0]{\begingroup\@sanitize@url \@url }%
\providecommand \@url [1]{\endgroup\@href {#1}{\urlprefix }}%
\providecommand \urlprefix  [0]{URL }%
\providecommand \Eprint [0]{\href }%
\providecommand \doibase [0]{https://doi.org/}%
\providecommand \selectlanguage [0]{\@gobble}%
\providecommand \bibinfo  [0]{\@secondoftwo}%
\providecommand \bibfield  [0]{\@secondoftwo}%
\providecommand \translation [1]{[#1]}%
\providecommand \BibitemOpen [0]{}%
\providecommand \bibitemStop [0]{}%
\providecommand \bibitemNoStop [0]{.\EOS\space}%
\providecommand \EOS [0]{\spacefactor3000\relax}%
\providecommand \BibitemShut  [1]{\csname bibitem#1\endcsname}%
\let\auto@bib@innerbib\@empty
%</preamble>
\bibitem [{\citenamefont {Budker}\ \emph {et~al.}(2002)\citenamefont {Budker},
  \citenamefont {Gawlik}, \citenamefont {Kimball}, \citenamefont {Rochester},
  \citenamefont {Yaschuk},\ and\ \citenamefont {Weis}}]{RevModPhys74.1153}%
  \BibitemOpen
  \bibfield  {author} {\bibinfo {author} {\bibfnamefont {D.}~\bibnamefont
  {Budker}}, \bibinfo {author} {\bibfnamefont {W.}~\bibnamefont {Gawlik}},
  \bibinfo {author} {\bibfnamefont {D.}~\bibnamefont {Kimball}}, \bibinfo
  {author} {\bibfnamefont {S.~R.}\ \bibnamefont {Rochester}}, \bibinfo {author}
  {\bibfnamefont {V.~V.}\ \bibnamefont {Yaschuk}},\ and\ \bibinfo {author}
  {\bibfnamefont {A.}~\bibnamefont {Weis}},\ }\bibfield  {title} {\bibinfo
  {title} {{Resonant nonlinear magneto-optical effects in atoms}},\ }\href@noop
  {} {\bibfield  {journal} {\bibinfo  {journal} {Reviews of Modern Physics}\
  }\textbf {\bibinfo {volume} {74}},\ \bibinfo {pages} {1153} (\bibinfo {year}
  {2002})}\BibitemShut {NoStop}%
\bibitem [{\citenamefont {Auzinsh}\ \emph {et~al.}(2010)\citenamefont
  {Auzinsh}, \citenamefont {Budker},\ and\ \citenamefont {Rochester}}]{OPA}%
  \BibitemOpen
  \bibfield  {author} {\bibinfo {author} {\bibfnamefont {M.}~\bibnamefont
  {Auzinsh}}, \bibinfo {author} {\bibfnamefont {D.}~\bibnamefont {Budker}},\
  and\ \bibinfo {author} {\bibfnamefont {S.~M.}\ \bibnamefont {Rochester}},\
  }\bibinfo {title} {Optically polarized atoms: Understanding light-atom
  interactions}\ (\bibinfo  {publisher} {Oxford University Press},\ \bibinfo
  {year} {2010})\BibitemShut {NoStop}%
\bibitem [{\citenamefont {Olsen}\ \emph {et~al.}(2011)\citenamefont {Olsen},
  \citenamefont {Patton}, \citenamefont {Jau},\ and\ \citenamefont
  {Happer}}]{PhysRevA84.063410}%
  \BibitemOpen
  \bibfield  {author} {\bibinfo {author} {\bibfnamefont {B.~A.}\ \bibnamefont
  {Olsen}}, \bibinfo {author} {\bibfnamefont {B.}~\bibnamefont {Patton}},
  \bibinfo {author} {\bibfnamefont {Y.~Y.}\ \bibnamefont {Jau}},\ and\ \bibinfo
  {author} {\bibfnamefont {W.}~\bibnamefont {Happer}},\ }\bibfield  {title}
  {\bibinfo {title} {{Optical pumping and spectroscopy of Cs vapor at high
  magnetic field}},\ }\href@noop {} {\bibfield  {journal} {\bibinfo  {journal}
  {Physical Review A}\ }\textbf {\bibinfo {volume} {84}},\ \bibinfo {pages}
  {063410} (\bibinfo {year} {2011})}\BibitemShut {NoStop}%
\bibitem [{\citenamefont {Klinger}\ \emph {et~al.}(2020)\citenamefont
  {Klinger}, \citenamefont {Azizbekyan}, \citenamefont {Sargsyan},
  \citenamefont {Leroy}, \citenamefont {Sarkisyan},\ and\ \citenamefont
  {Papoyan}}]{ApplOpt59.2231}%
  \BibitemOpen
  \bibfield  {author} {\bibinfo {author} {\bibfnamefont {E.}~\bibnamefont
  {Klinger}}, \bibinfo {author} {\bibfnamefont {H.}~\bibnamefont {Azizbekyan}},
  \bibinfo {author} {\bibfnamefont {A.}~\bibnamefont {Sargsyan}}, \bibinfo
  {author} {\bibfnamefont {C.}~\bibnamefont {Leroy}}, \bibinfo {author}
  {\bibfnamefont {D.}~\bibnamefont {Sarkisyan}},\ and\ \bibinfo {author}
  {\bibfnamefont {A.}~\bibnamefont {Papoyan}},\ }\bibfield  {title} {\bibinfo
  {title} {{Proof of the feasibility of a nanocell-based wide-range optical
  magnetometer}},\ }\href@noop {} {\bibfield  {journal} {\bibinfo  {journal}
  {Applied Optics}\ }\textbf {\bibinfo {volume} {59}},\ \bibinfo {pages} {2231}
  (\bibinfo {year} {2020})}\BibitemShut {NoStop}%
\bibitem [{\citenamefont {Zentile}\ \emph {et~al.}(2015)\citenamefont
  {Zentile}, \citenamefont {Keaveney}, \citenamefont {Weller}, \citenamefont
  {Whiting}, \citenamefont {Adams},\ and\ \citenamefont
  {Hughes}}]{CompPhysComm189.162}%
  \BibitemOpen
  \bibfield  {author} {\bibinfo {author} {\bibfnamefont {M.}~\bibnamefont
  {Zentile}}, \bibinfo {author} {\bibfnamefont {J.}~\bibnamefont {Keaveney}},
  \bibinfo {author} {\bibfnamefont {L.}~\bibnamefont {Weller}}, \bibinfo
  {author} {\bibfnamefont {D.~J.}\ \bibnamefont {Whiting}}, \bibinfo {author}
  {\bibfnamefont {C.~S.}\ \bibnamefont {Adams}},\ and\ \bibinfo {author}
  {\bibfnamefont {I.~G.}\ \bibnamefont {Hughes}},\ }\bibfield  {title}
  {\bibinfo {title} {{ElecSus: A program to calculate the electric
  susceptibility of an atomic ensemble}},\ }\href@noop {} {\bibfield  {journal}
  {\bibinfo  {journal} {Computer Physics Communications}\ }\textbf {\bibinfo
  {volume} {189}},\ \bibinfo {pages} {162} (\bibinfo {year}
  {2015})}\BibitemShut {NoStop}%
\bibitem [{\citenamefont {Tiesinga}\ \emph {et~al.}(2019)\citenamefont
  {Tiesinga}, \citenamefont {Mohr}, \citenamefont {Newell},\ and\ \citenamefont
  {Taylor}}]{CODATA}%
  \BibitemOpen
  \bibfield  {author} {\bibinfo {author} {\bibfnamefont {E.}~\bibnamefont
  {Tiesinga}}, \bibinfo {author} {\bibfnamefont {P.~J.}\ \bibnamefont {Mohr}},
  \bibinfo {author} {\bibfnamefont {D.~B.}\ \bibnamefont {Newell}},\ and\
  \bibinfo {author} {\bibfnamefont {B.~N.}\ \bibnamefont {Taylor}},\ }\bibfield
   {title} {\bibinfo {title} {The 2018 codata recommended values of the
  fundamental physical constants},\ }\href@noop {} {\bibfield  {journal}
  {\bibinfo  {journal} {NIST}\ } (\bibinfo {year} {2019})}\BibitemShut
  {NoStop}%
\bibitem [{\citenamefont {Arimondo}\ \emph {et~al.}(1977)\citenamefont
  {Arimondo}, \citenamefont {Inguscio},\ and\ \citenamefont
  {Violino}}]{RevModPhys.49.31}%
  \BibitemOpen
  \bibfield  {author} {\bibinfo {author} {\bibfnamefont {E.}~\bibnamefont
  {Arimondo}}, \bibinfo {author} {\bibfnamefont {M.}~\bibnamefont {Inguscio}},\
  and\ \bibinfo {author} {\bibfnamefont {P.}~\bibnamefont {Violino}},\
  }\bibfield  {title} {\bibinfo {title} {{Experimental determinations of the
  hyperfine structure in the alkali atoms}},\ }\href@noop {} {\bibfield
  {journal} {\bibinfo  {journal} {Reviews of Modern Physics}\ }\textbf
  {\bibinfo {volume} {49}},\ \bibinfo {pages} {31} (\bibinfo {year}
  {1977})}\BibitemShut {NoStop}%
\bibitem [{\citenamefont {Paschen}\ and\ \citenamefont {Back}(1921)}]{HPB}%
  \BibitemOpen
  \bibfield  {author} {\bibinfo {author} {\bibfnamefont {F.}~\bibnamefont
  {Paschen}}\ and\ \bibinfo {author} {\bibfnamefont {E.}~\bibnamefont {Back}},\
  }\bibinfo {title} {{Physica 1}}\ (\bibinfo {year} {1921})\BibitemShut
  {NoStop}%
\bibitem [{\citenamefont {Sargsyan}\ \emph {et~al.}(2012)\citenamefont
  {Sargsyan}, \citenamefont {Hahkumyan}, \citenamefont {Leroy}, \citenamefont
  {Pashayan-Leroy}, \citenamefont {Papoyan},\ and\ \citenamefont
  {Sarkisyan}}]{OptLett37.1379}%
  \BibitemOpen
  \bibfield  {author} {\bibinfo {author} {\bibfnamefont {A.}~\bibnamefont
  {Sargsyan}}, \bibinfo {author} {\bibfnamefont {G.}~\bibnamefont {Hahkumyan}},
  \bibinfo {author} {\bibfnamefont {C.}~\bibnamefont {Leroy}}, \bibinfo
  {author} {\bibfnamefont {Y.}~\bibnamefont {Pashayan-Leroy}}, \bibinfo
  {author} {\bibfnamefont {A.}~\bibnamefont {Papoyan}},\ and\ \bibinfo {author}
  {\bibfnamefont {D.}~\bibnamefont {Sarkisyan}},\ }\bibfield  {title} {\bibinfo
  {title} {{Hyperfine Paschen–Back regime realized in Rb nanocell}},\
  }\href@noop {} {\bibfield  {journal} {\bibinfo  {journal} {Optics Letters}\
  }\textbf {\bibinfo {volume} {37}},\ \bibinfo {pages} {1379} (\bibinfo {year}
  {2012})}\BibitemShut {NoStop}%
\bibitem [{\citenamefont {Weller}\ \emph {et~al.}(2012)\citenamefont {Weller},
  \citenamefont {Kleinbach}, \citenamefont {Zentile}, \citenamefont {Knappe},
  \citenamefont {Hughes},\ and\ \citenamefont {Adams}}]{OptLett37.3405}%
  \BibitemOpen
  \bibfield  {author} {\bibinfo {author} {\bibfnamefont {L.}~\bibnamefont
  {Weller}}, \bibinfo {author} {\bibfnamefont {K.~S.}\ \bibnamefont
  {Kleinbach}}, \bibinfo {author} {\bibfnamefont {M.~A.}\ \bibnamefont
  {Zentile}}, \bibinfo {author} {\bibfnamefont {S.}~\bibnamefont {Knappe}},
  \bibinfo {author} {\bibfnamefont {I.~G.}\ \bibnamefont {Hughes}},\ and\
  \bibinfo {author} {\bibfnamefont {C.~S.}\ \bibnamefont {Adams}},\ }\bibfield
  {title} {\bibinfo {title} {{Optical isolator using an atomic vapor in the
  hyperfine Paschen–Back regime}},\ }\href@noop {} {\bibfield  {journal}
  {\bibinfo  {journal} {Optics Letters}\ }\textbf {\bibinfo {volume} {37}},\
  \bibinfo {pages} {3405} (\bibinfo {year} {2012})}\BibitemShut {NoStop}%
\bibitem [{\citenamefont {Scotto}\ \emph {et~al.}(2015)\citenamefont {Scotto},
  \citenamefont {Ciampini}, \citenamefont {Rizzo},\ and\ \citenamefont
  {Arimondo}}]{PhysRevA92.063810}%
  \BibitemOpen
  \bibfield  {author} {\bibinfo {author} {\bibfnamefont {S.}~\bibnamefont
  {Scotto}}, \bibinfo {author} {\bibfnamefont {D.}~\bibnamefont {Ciampini}},
  \bibinfo {author} {\bibfnamefont {C.}~\bibnamefont {Rizzo}},\ and\ \bibinfo
  {author} {\bibfnamefont {E.}~\bibnamefont {Arimondo}},\ }\bibfield  {title}
  {\bibinfo {title} {{Four-level N-scheme crossover resonances in Rb saturation
  spectroscopy in magnetic fields}},\ }\href@noop {} {\bibfield  {journal}
  {\bibinfo  {journal} {Physical Review A}\ }\textbf {\bibinfo {volume} {92}},\
  \bibinfo {pages} {063810} (\bibinfo {year} {2015})}\BibitemShut {NoStop}%
\bibitem [{\citenamefont {Whiting}\ \emph {et~al.}(2018)\citenamefont
  {Whiting}, \citenamefont {Matthew}, \citenamefont {Keaveney}, \citenamefont
  {Adams},\ and\ \citenamefont {Hughes}}]{JModOpt65.713}%
  \BibitemOpen
  \bibfield  {author} {\bibinfo {author} {\bibfnamefont {D.~J.}\ \bibnamefont
  {Whiting}}, \bibinfo {author} {\bibfnamefont {R.~S.}\ \bibnamefont
  {Matthew}}, \bibinfo {author} {\bibfnamefont {J.}~\bibnamefont {Keaveney}},
  \bibinfo {author} {\bibfnamefont {C.~S.}\ \bibnamefont {Adams}},\ and\
  \bibinfo {author} {\bibfnamefont {I.~G.}\ \bibnamefont {Hughes}},\ }\bibfield
   {title} {\bibinfo {title} {{Four-wave mixing in a non-degenerate four-level
  diamond configuration in the hyperfine Paschen–Back regime}},\ }\href@noop
  {} {\bibfield  {journal} {\bibinfo  {journal} {Journal of Modern Optics}\
  }\textbf {\bibinfo {volume} {65}},\ \bibinfo {pages} {713} (\bibinfo {year}
  {2018})}\BibitemShut {NoStop}%
\bibitem [{\citenamefont {Hakhumyan}\ \emph {et~al.}(2011)\citenamefont
  {Hakhumyan}, \citenamefont {Leroy}, \citenamefont {Pashayan-Leroy},
  \citenamefont {Sarkisyan},\ and\ \citenamefont {Auzinsh}}]{OptComm284.4007}%
  \BibitemOpen
  \bibfield  {author} {\bibinfo {author} {\bibfnamefont {G.}~\bibnamefont
  {Hakhumyan}}, \bibinfo {author} {\bibfnamefont {C.}~\bibnamefont {Leroy}},
  \bibinfo {author} {\bibfnamefont {Y.}~\bibnamefont {Pashayan-Leroy}},
  \bibinfo {author} {\bibfnamefont {D.}~\bibnamefont {Sarkisyan}},\ and\
  \bibinfo {author} {\bibfnamefont {M.}~\bibnamefont {Auzinsh}},\ }\bibfield
  {title} {\bibinfo {title} {{High-spatial-resolution monitoring of strong
  magnetic field using Rb vapor nanometric-thin cell}},\ }\href@noop {}
  {\bibfield  {journal} {\bibinfo  {journal} {Optics Communications}\ }\textbf
  {\bibinfo {volume} {284}},\ \bibinfo {pages} {4007} (\bibinfo {year}
  {2011})}\BibitemShut {NoStop}%
\bibitem [{\citenamefont {Sargsyan}\ \emph {et~al.}(2014)\citenamefont
  {Sargsyan}, \citenamefont {Tonoyan}, \citenamefont {Mirzoyan}, \citenamefont
  {Sarkisyan}, \citenamefont {Wojciechowski},\ and\ \citenamefont
  {Gawlik}}]{OptLett39.2270}%
  \BibitemOpen
  \bibfield  {author} {\bibinfo {author} {\bibfnamefont {A.}~\bibnamefont
  {Sargsyan}}, \bibinfo {author} {\bibfnamefont {A.}~\bibnamefont {Tonoyan}},
  \bibinfo {author} {\bibfnamefont {R.}~\bibnamefont {Mirzoyan}}, \bibinfo
  {author} {\bibfnamefont {D.}~\bibnamefont {Sarkisyan}}, \bibinfo {author}
  {\bibfnamefont {A.}~\bibnamefont {Wojciechowski}},\ and\ \bibinfo {author}
  {\bibfnamefont {W.}~\bibnamefont {Gawlik}},\ }\bibfield  {title} {\bibinfo
  {title} {{Saturated-absorption spectroscopy revisited: atomic transitions in
  strong magnetic fields ($>$20 mT) with a micrometer-thin cell}},\ }\href@noop
  {} {\bibfield  {journal} {\bibinfo  {journal} {Optics Letters}\ }\textbf
  {\bibinfo {volume} {39}},\ \bibinfo {pages} {2270} (\bibinfo {year}
  {2014})}\BibitemShut {NoStop}%
\bibitem [{\citenamefont {Sargsyan}\ \emph {et~al.}(2017)\citenamefont
  {Sargsyan}, \citenamefont {Papoyan}, \citenamefont {Hughes}, \citenamefont
  {Adams},\ and\ \citenamefont {Sarkisyan}}]{OptLett42.1476}%
  \BibitemOpen
  \bibfield  {author} {\bibinfo {author} {\bibfnamefont {A.}~\bibnamefont
  {Sargsyan}}, \bibinfo {author} {\bibfnamefont {A.}~\bibnamefont {Papoyan}},
  \bibinfo {author} {\bibfnamefont {I.~G.}\ \bibnamefont {Hughes}}, \bibinfo
  {author} {\bibfnamefont {C.~S.}\ \bibnamefont {Adams}},\ and\ \bibinfo
  {author} {\bibfnamefont {D.}~\bibnamefont {Sarkisyan}},\ }\bibfield  {title}
  {\bibinfo {title} {{Selective reflection from an Rb layer with a thickness
  below $\lambda$/12 and applications}},\ }\href@noop {} {\bibfield  {journal}
  {\bibinfo  {journal} {Optics Letters}\ }\textbf {\bibinfo {volume} {42}},\
  \bibinfo {pages} {1476} (\bibinfo {year} {2017})}\BibitemShut {NoStop}%
\bibitem [{\citenamefont {Sargsyan}\ \emph {et~al.}(2015)\citenamefont
  {Sargsyan}, \citenamefont {Tonoyan}, \citenamefont {Hakhumyan}, \citenamefont
  {Leroy}, \citenamefont {Pashayan-Leroy},\ and\ \citenamefont
  {Sarkisyan}}]{EPL110.23001}%
  \BibitemOpen
  \bibfield  {author} {\bibinfo {author} {\bibfnamefont {A.}~\bibnamefont
  {Sargsyan}}, \bibinfo {author} {\bibfnamefont {A.}~\bibnamefont {Tonoyan}},
  \bibinfo {author} {\bibfnamefont {G.}~\bibnamefont {Hakhumyan}}, \bibinfo
  {author} {\bibfnamefont {C.}~\bibnamefont {Leroy}}, \bibinfo {author}
  {\bibfnamefont {Y.}~\bibnamefont {Pashayan-Leroy}},\ and\ \bibinfo {author}
  {\bibfnamefont {D.}~\bibnamefont {Sarkisyan}},\ }\bibfield  {title} {\bibinfo
  {title} {{Complete hyperfine Paschen-Back regime at relatively small magnetic
  fields realized in potassium nano-cell}},\ }\href@noop {} {\bibfield
  {journal} {\bibinfo  {journal} {Europhysics Letters}\ }\textbf {\bibinfo
  {volume} {110}},\ \bibinfo {pages} {23001} (\bibinfo {year}
  {2015})}\BibitemShut {NoStop}%
\bibitem [{\citenamefont {Sargsyan}\ \emph {et~al.}(2018)\citenamefont
  {Sargsyan}, \citenamefont {Tonoyan}, \citenamefont {Keaveney}, \citenamefont
  {Hughes}, \citenamefont {Adams},\ and\ \citenamefont
  {Sarkisyan}}]{JETP153.355}%
  \BibitemOpen
  \bibfield  {author} {\bibinfo {author} {\bibfnamefont {A.}~\bibnamefont
  {Sargsyan}}, \bibinfo {author} {\bibfnamefont {A.}~\bibnamefont {Tonoyan}},
  \bibinfo {author} {\bibfnamefont {J.}~\bibnamefont {Keaveney}}, \bibinfo
  {author} {\bibfnamefont {I.~G.}\ \bibnamefont {Hughes}}, \bibinfo {author}
  {\bibfnamefont {C.~S.}\ \bibnamefont {Adams}},\ and\ \bibinfo {author}
  {\bibfnamefont {D.}~\bibnamefont {Sarkisyan}},\ }\bibfield  {title} {\bibinfo
  {title} {{Selective Reflection of Potassium Vapor Nanolayers in a Magnetic
  Field}},\ }\href@noop {} {\bibfield  {journal} {\bibinfo  {journal} {Journal
  of Experimental and Theoretical Physics}\ }\textbf {\bibinfo {volume}
  {153}},\ \bibinfo {pages} {355} (\bibinfo {year} {2018})}\BibitemShut
  {NoStop}%
\bibitem [{\citenamefont {Sargsyan}\ \emph {et~al.}(2016)\citenamefont
  {Sargsyan}, \citenamefont {Klinger}, \citenamefont {Pashayan-Leroy},
  \citenamefont {Leroy}, \citenamefont {Papoyan},\ and\ \citenamefont
  {Sarkisyan}}]{JETPLett104.224}%
  \BibitemOpen
  \bibfield  {author} {\bibinfo {author} {\bibfnamefont {A.}~\bibnamefont
  {Sargsyan}}, \bibinfo {author} {\bibfnamefont {E.}~\bibnamefont {Klinger}},
  \bibinfo {author} {\bibfnamefont {Y.}~\bibnamefont {Pashayan-Leroy}},
  \bibinfo {author} {\bibfnamefont {C.}~\bibnamefont {Leroy}}, \bibinfo
  {author} {\bibfnamefont {A.}~\bibnamefont {Papoyan}},\ and\ \bibinfo {author}
  {\bibfnamefont {D.}~\bibnamefont {Sarkisyan}},\ }\bibfield  {title} {\bibinfo
  {title} {{Selective reflection from Rb vapor in half- and quarter-wave cells:
  Features and possible applications}},\ }\href@noop {} {\bibfield  {journal}
  {\bibinfo  {journal} {Journal of Experimental and Theoretical Physics
  Letters}\ }\textbf {\bibinfo {volume} {104}},\ \bibinfo {pages} {224}
  (\bibinfo {year} {2016})}\BibitemShut {NoStop}%
\bibitem [{\citenamefont {Dutier}\ \emph {et~al.}(2003)\citenamefont {Dutier},
  \citenamefont {Saltiel}, \citenamefont {Bloch},\ and\ \citenamefont
  {Ducloy}}]{JOSAB20.793}%
  \BibitemOpen
  \bibfield  {author} {\bibinfo {author} {\bibfnamefont {G.}~\bibnamefont
  {Dutier}}, \bibinfo {author} {\bibfnamefont {S.}~\bibnamefont {Saltiel}},
  \bibinfo {author} {\bibfnamefont {D.}~\bibnamefont {Bloch}},\ and\ \bibinfo
  {author} {\bibfnamefont {M.}~\bibnamefont {Ducloy}},\ }\bibfield  {title}
  {\bibinfo {title} {{Revisiting optical spectroscopy in a thin vapor cell:
  mixing of reflection and transmission as a Fabry–Perot microcavity
  effect}},\ }\href@noop {} {\bibfield  {journal} {\bibinfo  {journal} {Journal
  of the Optical Society of America B}\ }\textbf {\bibinfo {volume} {20}},\
  \bibinfo {pages} {793} (\bibinfo {year} {2003})}\BibitemShut {NoStop}%
\bibitem [{\citenamefont {Vartanyan}\ and\ \citenamefont
  {Lin}(1995)}]{PhysRevA.51.1959}%
  \BibitemOpen
  \bibfield  {author} {\bibinfo {author} {\bibfnamefont {T.~A.}\ \bibnamefont
  {Vartanyan}}\ and\ \bibinfo {author} {\bibfnamefont {D.~L.}\ \bibnamefont
  {Lin}},\ }\bibfield  {title} {\bibinfo {title} {{Enhanced selective
  reflection from a thin layer of a dilute gaseous medium}},\ }\href@noop {}
  {\bibfield  {journal} {\bibinfo  {journal} {Physical Review A}\ }\textbf
  {\bibinfo {volume} {51}},\ \bibinfo {pages} {1959} (\bibinfo {year}
  {1995})}\BibitemShut {NoStop}%
\bibitem [{\citenamefont {Tremblay}\ \emph {et~al.}(1990)\citenamefont
  {Tremblay}, \citenamefont {Michaud}, \citenamefont {Levesque}, \citenamefont
  {Theriault}, \citenamefont {Breton}, \citenamefont {Beaubien},\ and\
  \citenamefont {Cyr}}]{PhysRevA42.2766}%
  \BibitemOpen
  \bibfield  {author} {\bibinfo {author} {\bibfnamefont {P.}~\bibnamefont
  {Tremblay}}, \bibinfo {author} {\bibfnamefont {A.}~\bibnamefont {Michaud}},
  \bibinfo {author} {\bibfnamefont {M.}~\bibnamefont {Levesque}}, \bibinfo
  {author} {\bibfnamefont {S.}~\bibnamefont {Theriault}}, \bibinfo {author}
  {\bibfnamefont {M.}~\bibnamefont {Breton}}, \bibinfo {author} {\bibfnamefont
  {J.}~\bibnamefont {Beaubien}},\ and\ \bibinfo {author} {\bibfnamefont
  {N.}~\bibnamefont {Cyr}},\ }\bibfield  {title} {\bibinfo {title} {{Absorption
  profiles of alkali-metal $D$ lines in the presence of a static magnetic
  field}},\ }\href@noop {} {\bibfield  {journal} {\bibinfo  {journal} {Physical
  Review A}\ }\textbf {\bibinfo {volume} {42}},\ \bibinfo {pages} {2766}
  (\bibinfo {year} {1990})}\BibitemShut {NoStop}%
\bibitem [{\citenamefont {Momier}\ \emph {et~al.}(2021)\citenamefont {Momier},
  \citenamefont {Papoyan},\ and\ \citenamefont {Leroy}}]{JQSRT272.107780}%
  \BibitemOpen
  \bibfield  {author} {\bibinfo {author} {\bibfnamefont {R.}~\bibnamefont
  {Momier}}, \bibinfo {author} {\bibfnamefont {A.}~\bibnamefont {Papoyan}},\
  and\ \bibinfo {author} {\bibfnamefont {C.}~\bibnamefont {Leroy}},\ }\bibfield
   {title} {\bibinfo {title} {{Sub-Doppler spectra of sodium $D$ lines in a
  wide range of magnetic field: Theoretical study}},\ }\href@noop {} {\bibfield
   {journal} {\bibinfo  {journal} {Journal of Quantitative Spectroscopy and
  Radiative Transfer}\ }\textbf {\bibinfo {volume} {272}},\ \bibinfo {pages}
  {107780} (\bibinfo {year} {2021})}\BibitemShut {NoStop}%
\bibitem [{\citenamefont {Vassiliev}\ \emph {et~al.}(2006)\citenamefont
  {Vassiliev}, \citenamefont {Zibrov},\ and\ \citenamefont
  {Velichansky}}]{RevSciInstrum77.013102}%
  \BibitemOpen
  \bibfield  {author} {\bibinfo {author} {\bibfnamefont {V.~V.}\ \bibnamefont
  {Vassiliev}}, \bibinfo {author} {\bibfnamefont {S.~A.}\ \bibnamefont
  {Zibrov}},\ and\ \bibinfo {author} {\bibfnamefont {V.~L.}\ \bibnamefont
  {Velichansky}},\ }\bibfield  {title} {\bibinfo {title} {{Compact
  extended-cavity diode laser for atomic spectroscopy and metrology}},\
  }\href@noop {} {\bibfield  {journal} {\bibinfo  {journal} {Review of
  Scientific Instruments}\ }\textbf {\bibinfo {volume} {77}},\ \bibinfo {pages}
  {013102} (\bibinfo {year} {2006})}\BibitemShut {NoStop}%
\bibitem [{\citenamefont {Bloch}\ \emph {et~al.}(1996)\citenamefont {Bloch},
  \citenamefont {Ducloy}, \citenamefont {Senkov}, \citenamefont {Velichansky},\
  and\ \citenamefont {Yudin}}]{LasPhys6.670}%
  \BibitemOpen
  \bibfield  {author} {\bibinfo {author} {\bibfnamefont {D.}~\bibnamefont
  {Bloch}}, \bibinfo {author} {\bibfnamefont {M.}~\bibnamefont {Ducloy}},
  \bibinfo {author} {\bibfnamefont {N.}~\bibnamefont {Senkov}}, \bibinfo
  {author} {\bibfnamefont {V.}~\bibnamefont {Velichansky}},\ and\ \bibinfo
  {author} {\bibfnamefont {V.}~\bibnamefont {Yudin}},\ }\bibfield  {title}
  {\bibinfo {title} {{Doppler-Free Spectroscopy of the D1 Line of Potassium}},\
  }\href@noop {} {\bibfield  {journal} {\bibinfo  {journal} {Laser Physics}\
  }\textbf {\bibinfo {volume} {6}},\ \bibinfo {pages} {670} (\bibinfo {year}
  {1996})}\BibitemShut {NoStop}%
\bibitem [{\citenamefont {Nakayama}(1985)}]{JOSAB2.1431}%
  \BibitemOpen
  \bibfield  {author} {\bibinfo {author} {\bibfnamefont {S.}~\bibnamefont
  {Nakayama}},\ }\bibfield  {title} {\bibinfo {title} {{Doppler-free laser
  spectroscopic techniques with optical pumping in $D_1$ lines of alkali
  atoms}},\ }\href@noop {} {\bibfield  {journal} {\bibinfo  {journal} {Journal
  of the Optical Society of America B}\ }\textbf {\bibinfo {volume} {2}},\
  \bibinfo {pages} {1431} (\bibinfo {year} {1985})}\BibitemShut {NoStop}%
\bibitem [{\citenamefont {Preston}(1996)}]{AJP64.1432}%
  \BibitemOpen
  \bibfield  {author} {\bibinfo {author} {\bibfnamefont {D.~W.}\ \bibnamefont
  {Preston}},\ }\bibfield  {title} {\bibinfo {title} {{Doppler‐free saturated
  absorption: Laser spectroscopy}},\ }\href@noop {} {\bibfield  {journal}
  {\bibinfo  {journal} {American Journal of Physics}\ }\textbf {\bibinfo
  {volume} {64}},\ \bibinfo {pages} {1432} (\bibinfo {year}
  {1996})}\BibitemShut {NoStop}%
\bibitem [{\citenamefont {Jahier}\ \emph {et~al.}(2000)\citenamefont {Jahier},
  \citenamefont {Guéna}, \citenamefont {Jacquier}, \citenamefont {Lintz},
  \citenamefont {Papoyan},\ and\ \citenamefont {Bouchiat}}]{AppPhysB71.561}%
  \BibitemOpen
  \bibfield  {author} {\bibinfo {author} {\bibfnamefont {E.}~\bibnamefont
  {Jahier}}, \bibinfo {author} {\bibfnamefont {J.}~\bibnamefont {Guéna}},
  \bibinfo {author} {\bibfnamefont {P.}~\bibnamefont {Jacquier}}, \bibinfo
  {author} {\bibfnamefont {M.}~\bibnamefont {Lintz}}, \bibinfo {author}
  {\bibfnamefont {A.~V.}\ \bibnamefont {Papoyan}},\ and\ \bibinfo {author}
  {\bibfnamefont {M.~A.}\ \bibnamefont {Bouchiat}},\ }\bibfield  {title}
  {\bibinfo {title} {{Temperature-tunable sapphire windows for reflection
  loss-free operation of vapor cells}},\ }\href@noop {} {\bibfield  {journal}
  {\bibinfo  {journal} {Applied Physics B}\ }\textbf {\bibinfo {volume} {71}},\
  \bibinfo {pages} {561} (\bibinfo {year} {2000})}\BibitemShut {NoStop}%
\bibitem [{\citenamefont {Silvfast}(2004)}]{LasFun}%
  \BibitemOpen
  \bibfield  {author} {\bibinfo {author} {\bibfnamefont {W.~T.}\ \bibnamefont
  {Silvfast}},\ }\bibinfo {title} {{Laser Fundamentals}}\ (\bibinfo
  {publisher} {Cambridge University Press},\ \bibinfo {year} {2004})\ \bibinfo
  {edition} {2nd}\ ed.\BibitemShut {Stop}%
\bibitem [{\citenamefont {Sargsyan}\ \emph {et~al.}(2019)\citenamefont
  {Sargsyan}, \citenamefont {Amiryan}, \citenamefont {Pashayan-Leroy},
  \citenamefont {Leroy}, \citenamefont {Papoyan},\ and\ \citenamefont
  {Sarkisyan}}]{OptLett44.5533}%
  \BibitemOpen
  \bibfield  {author} {\bibinfo {author} {\bibfnamefont {A.}~\bibnamefont
  {Sargsyan}}, \bibinfo {author} {\bibfnamefont {A.}~\bibnamefont {Amiryan}},
  \bibinfo {author} {\bibfnamefont {Y.}~\bibnamefont {Pashayan-Leroy}},
  \bibinfo {author} {\bibfnamefont {C.}~\bibnamefont {Leroy}}, \bibinfo
  {author} {\bibfnamefont {A.}~\bibnamefont {Papoyan}},\ and\ \bibinfo {author}
  {\bibfnamefont {D.}~\bibnamefont {Sarkisyan}},\ }\bibfield  {title} {\bibinfo
  {title} {{Approach to quantitative spectroscopy of atomic vapor in optical
  nanocells }},\ }\href@noop {} {\bibfield  {journal} {\bibinfo  {journal}
  {Optics Letters}\ }\textbf {\bibinfo {volume} {44}},\ \bibinfo {pages} {5533}
  (\bibinfo {year} {2019})}\BibitemShut {NoStop}%
\bibitem [{\citenamefont {Romer}\ and\ \citenamefont
  {Dicke}(1955)}]{PhysRev.99.532}%
  \BibitemOpen
  \bibfield  {author} {\bibinfo {author} {\bibfnamefont {R.~H.}\ \bibnamefont
  {Romer}}\ and\ \bibinfo {author} {\bibfnamefont {R.~H.}\ \bibnamefont
  {Dicke}},\ }\bibfield  {title} {\bibinfo {title} {{New Technique for
  High-Resolution Microwave Spectroscopy}},\ }\href
  {https://doi.org/10.1103/PhysRev.99.532} {\bibfield  {journal} {\bibinfo
  {journal} {Physical Review}\ }\textbf {\bibinfo {volume} {99}},\ \bibinfo
  {pages} {532} (\bibinfo {year} {1955})}\BibitemShut {NoStop}%
\bibitem [{\citenamefont {Sargsyan}\ \emph {et~al.}(2021)\citenamefont
  {Sargsyan}, \citenamefont {Klinger}, \citenamefont {Amiryan}, \citenamefont
  {Tonoyan},\ and\ \citenamefont {Sarkisyan}}]{PhysLettA390.127114}%
  \BibitemOpen
  \bibfield  {author} {\bibinfo {author} {\bibfnamefont {A.}~\bibnamefont
  {Sargsyan}}, \bibinfo {author} {\bibfnamefont {E.}~\bibnamefont {Klinger}},
  \bibinfo {author} {\bibfnamefont {A.}~\bibnamefont {Amiryan}}, \bibinfo
  {author} {\bibfnamefont {A.}~\bibnamefont {Tonoyan}},\ and\ \bibinfo {author}
  {\bibfnamefont {D.}~\bibnamefont {Sarkisyan}},\ }\bibfield  {title} {\bibinfo
  {title} {{Circular dichroism in atomic vapors: Magnetically induced
  transitions responsible for two distinct behaviors}},\ }\href@noop {}
  {\bibfield  {journal} {\bibinfo  {journal} {Physics Letters A}\ }\textbf
  {\bibinfo {volume} {390}},\ \bibinfo {pages} {127114} (\bibinfo {year}
  {2021})}\BibitemShut {NoStop}%
\bibitem [{\citenamefont {Sargsyan}\ \emph
  {et~al.}(2022{\natexlab{a}})\citenamefont {Sargsyan}, \citenamefont
  {Tonoyan}, \citenamefont {Momier}, \citenamefont {Leroy},\ and\ \citenamefont
  {Sarkisyan}}]{JOSAB39.973}%
  \BibitemOpen
  \bibfield  {author} {\bibinfo {author} {\bibfnamefont {A.}~\bibnamefont
  {Sargsyan}}, \bibinfo {author} {\bibfnamefont {A.}~\bibnamefont {Tonoyan}},
  \bibinfo {author} {\bibfnamefont {R.}~\bibnamefont {Momier}}, \bibinfo
  {author} {\bibfnamefont {C.}~\bibnamefont {Leroy}},\ and\ \bibinfo {author}
  {\bibfnamefont {D.}~\bibnamefont {Sarkisyan}},\ }\bibfield  {title} {\bibinfo
  {title} {{Dominant magnetically induced transitions in alkali metal atoms
  with nuclear spin 3/2 }},\ }\href@noop {} {\bibfield  {journal} {\bibinfo
  {journal} {Journal of the Optical Society of America B}\ }\textbf {\bibinfo
  {volume} {39}},\ \bibinfo {pages} {973} (\bibinfo {year}
  {2022}{\natexlab{a}})}\BibitemShut {NoStop}%
\bibitem [{\citenamefont {Sargsyan}\ \emph
  {et~al.}(2022{\natexlab{b}})\citenamefont {Sargsyan}, \citenamefont
  {Amiryan}, \citenamefont {Tonoyan}, \citenamefont {Klinger},\ and\
  \citenamefont {Sarkisyan}}]{PhysLettA434.128043}%
  \BibitemOpen
  \bibfield  {author} {\bibinfo {author} {\bibfnamefont {A.}~\bibnamefont
  {Sargsyan}}, \bibinfo {author} {\bibfnamefont {A.}~\bibnamefont {Amiryan}},
  \bibinfo {author} {\bibfnamefont {A.}~\bibnamefont {Tonoyan}}, \bibinfo
  {author} {\bibfnamefont {E.}~\bibnamefont {Klinger}},\ and\ \bibinfo {author}
  {\bibfnamefont {D.}~\bibnamefont {Sarkisyan}},\ }\bibfield  {title} {\bibinfo
  {title} {{Coherent optical processes on Cs $D_2$ line magnetically induced
  transitions}},\ }\href@noop {} {\bibfield  {journal} {\bibinfo  {journal}
  {Physics Letters A}\ }\textbf {\bibinfo {volume} {434}},\ \bibinfo {pages}
  {128043} (\bibinfo {year} {2022}{\natexlab{b}})}\BibitemShut {NoStop}%
\bibitem [{\citenamefont {Krasteva}\ \emph {et~al.}(2020)\citenamefont
  {Krasteva}, \citenamefont {Gosh}, \citenamefont {Gateva}, \citenamefont
  {Tsvetkov}, \citenamefont {Sarkisyan}, \citenamefont {Sargsyan},
  \citenamefont {Vartanyan},\ and\ \citenamefont
  {Cartaleva}}]{PhysScript95.015404}%
  \BibitemOpen
  \bibfield  {author} {\bibinfo {author} {\bibfnamefont {A.}~\bibnamefont
  {Krasteva}}, \bibinfo {author} {\bibfnamefont {P.}~\bibnamefont {Gosh}},
  \bibinfo {author} {\bibfnamefont {S.}~\bibnamefont {Gateva}}, \bibinfo
  {author} {\bibfnamefont {S.}~\bibnamefont {Tsvetkov}}, \bibinfo {author}
  {\bibfnamefont {D.}~\bibnamefont {Sarkisyan}}, \bibinfo {author}
  {\bibfnamefont {A.}~\bibnamefont {Sargsyan}}, \bibinfo {author}
  {\bibfnamefont {T.}~\bibnamefont {Vartanyan}},\ and\ \bibinfo {author}
  {\bibfnamefont {S.}~\bibnamefont {Cartaleva}},\ }\bibfield  {title} {\bibinfo
  {title} {{Observation and theoretical simulation of $N$-resonances in Cs
  $D_2$ lines}},\ }\href@noop {} {\bibfield  {journal} {\bibinfo  {journal}
  {Physica Scripta}\ }\textbf {\bibinfo {volume} {95}},\ \bibinfo {pages}
  {015404} (\bibinfo {year} {2020})}\BibitemShut {NoStop}%
\bibitem [{BLO(2005)}]{BLOCH200591}%
  \BibitemOpen
  \bibfield  {title} {\bibinfo {title} {{Atom-wall interaction}}\ }(\bibinfo
  {publisher} {Academic Press},\ \bibinfo {year} {2005})\ pp.\ \bibinfo {pages}
  {91--154}\BibitemShut {NoStop}%
\bibitem [{\citenamefont {Whittaker}\ \emph {et~al.}(2014)\citenamefont
  {Whittaker}, \citenamefont {Keaveney}, \citenamefont {Hughes}, \citenamefont
  {Sargsyan}, \citenamefont {Sarkisyan},\ and\ \citenamefont
  {Adams}}]{PhysRevLett.112.253201}%
  \BibitemOpen
  \bibfield  {author} {\bibinfo {author} {\bibfnamefont {K.~A.}\ \bibnamefont
  {Whittaker}}, \bibinfo {author} {\bibfnamefont {J.}~\bibnamefont {Keaveney}},
  \bibinfo {author} {\bibfnamefont {I.~G.}\ \bibnamefont {Hughes}}, \bibinfo
  {author} {\bibfnamefont {A.}~\bibnamefont {Sargsyan}}, \bibinfo {author}
  {\bibfnamefont {D.}~\bibnamefont {Sarkisyan}},\ and\ \bibinfo {author}
  {\bibfnamefont {C.~S.}\ \bibnamefont {Adams}},\ }\bibfield  {title} {\bibinfo
  {title} {{Optical Response of Gas-Phase Atoms at Less than
  $\ensuremath{\lambda}/80$ from a Dielectric Surface}},\ }\href@noop {}
  {\bibfield  {journal} {\bibinfo  {journal} {Phys. Rev. Lett.}\ }\textbf
  {\bibinfo {volume} {112}},\ \bibinfo {pages} {253201} (\bibinfo {year}
  {2014})}\BibitemShut {NoStop}%
\bibitem [{\citenamefont {Liang}\ \emph {et~al.}(2020)\citenamefont {Liang},
  \citenamefont {Fuchs}, \citenamefont {Schäfer},\ and\ \citenamefont
  {Kresin}}]{RevSciInstrum91.053202}%
  \BibitemOpen
  \bibfield  {author} {\bibinfo {author} {\bibfnamefont {J.}~\bibnamefont
  {Liang}}, \bibinfo {author} {\bibfnamefont {T.~M.}\ \bibnamefont {Fuchs}},
  \bibinfo {author} {\bibfnamefont {R.}~\bibnamefont {Schäfer}},\ and\
  \bibinfo {author} {\bibfnamefont {V.~V.}\ \bibnamefont {Kresin}},\ }\bibfield
   {title} {\bibinfo {title} {{Strong permanent magnet gradient deflector for
  Stern–Gerlach-type experiments on molecular beams}},\ }\href@noop {}
  {\bibfield  {journal} {\bibinfo  {journal} {Review of Scientific
  Instruments}\ }\textbf {\bibinfo {volume} {91}},\ \bibinfo {pages} {053202}
  (\bibinfo {year} {2020})}\BibitemShut {NoStop}%
\end{thebibliography}%

\end{document}